\documentclass[twocolumn]{emulateapj}
\journalinfo{{\sc Accepted by The Astrophysical Journal}}
\usepackage{natbib}
\usepackage{amssymb}
\usepackage{graphicx}

\newcommand{\chan}{\emph {Chandra}}

\begin{document}
\title{Geometrical Distance Determination using Type I X-ray Bursts}
  
\author{Thomas W. J. Thompson\altaffilmark{1},
Richard E. Rothschild\altaffilmark{1},
John A. Tomsick\altaffilmark{1}
}

\altaffiltext{1}{Center for Astrophysics and Space Sciences, University
of California, San Diego, La Jolla, CA 92093; tthompson@physics.ucsd.edu }

\begin{abstract}
With the excellent angular resolution of the {\em Chandra X-ray Observatory}, it is possible to geometrically determine the distance to variable Galactic sources, based on the phenomenon that scattered radiation appearing in the X-ray halo has to travel along a slightly longer path than the direct, unscattered radiation. By measuring the delayed variability, constraints on the source distance can be obtained if the halo brightness is large enough to dominate the point spread function (PSF) and to provide sufficient statistics. The distance to Cyg X-3, which has a quasi-sinusoidal light curve, has been obtained with this approach by Predehl et al. Here we examine the feasibility of using the delayed signature of type I X-ray bursts as distance indicators. We use simulations of delayed X-ray burst light curves in the halo to find that the optimal annular region and energy band for a distance measurement with a grating observation is roughly 10--50\arcsec~and 1--5 keV respectively, assuming \chan's effective area and PSF, uniformly distributed dust, the input spectrum and optical depth to GX 13+1, and the Weingartner \& Draine interstellar grain model. We find that the statistics are dominated by Poisson noise rather than systematic uncertainties, e.g., the PSF contribution to the halo. Using {\em Chandra}, a distance measurement to such a source at 4 (8) kpc could be made to about 23\% (30\%) accuracy with a single burst with 68\% confidence. By stacking many bursts, a reasonable estimate of systematic errors limit the distance measurement to about 10\% accuracy.
\end{abstract}

\section{Introduction}
Type I X-ray bursts potentially provide a delayed signature in the halo that can be used to measure the distance to variable X-ray sources with a method proposed by Tr\"{u}mper \& Sch\"{o}nfelder (1973; hereafter TS73), prior to the detection of the first X-ray halo around GX 339-4 by {\em Einstein} (Rolf 1983). The TS73 method is based on the phenomenon that grain-scattered photons from an X-ray source travel along slightly longer paths to the telescope than the direct, unscattered photons. Temporal variations in the intensity are therefore delayed and smeared when they appear in the X-ray halo. Using the single scattering approximation, it is possible to predict the delayed light curve as a function of halo angle, energy, and distance to an accuracy that is limited by knowledge of the distribution of dust along the line of sight and by the accuracy of the interstellar grain model. Hu et al. (2004) developed a similar method to measure distances with X-ray halo variability using the frequency domain. 

The feasibility of the TS73 distance measurement method has been demonstrated by Predehl et al. (2000), who derived a distance of 9$^{+4}_{-2}$ kpc to Cyg X-3 using the source's quasi-sinusoidal variability. In this paper, we examine the applicability of the TS73 method to the abrupt increase in flux provided by type I X-ray bursts, which result from unstable thermonuclear ignition of accreted material on the surface of the neutron star. Sharp and temporary intensity changes, such as that produced by type I bursts or gamma-ray bursts (GRBs), produce a qualitatively different response in the X-ray halo than the response from smooth intensity variations like those of Cyg X-3. For example, consider a smooth change in point source flux over an hour or so. Eventually the halo brightness will reach a new base level corresponding to the new point source intensity, and information on the source distance and dust distribution is contained in the rate of response and the resulting phase shift of the light curve. With X-ray bursts a new base level is not obtained; rather, the burst appears as a slight and temporary increase in halo flux after a time delay that depends on the scattering geometry. The intrinsically small signal-to-noise ($S/N$) ratio provided by bursts makes their potential application as distance indicators particularly difficult, requiring the physical parameters governing the delayed light curve in the X-ray halo be measured to a high degree of accuracy, and requiring careful treatment of the point spread function (PSF). 

Gamma-ray bursts produce a similar halo light curve, although the problem is much simpler: With spectroscopic determination of the distances to host galaxies of GRBs, the degeneracy between the distribution of dust and the distance to the source is removed. This has been exploited by both the {\em Swift} and {\em XMM-Newton} observatories to determine the distances to Galactic interstellar dust clouds along various sightlines (Vaughan et al. 2004; Tiengo \& Mereghetti 2006; Vaughan et al. 2006). 

The physical parameters governing the delayed light curve in the halo are the peak and persistent flux, the spectral shape, the scattering optical depth, the scale time of the approximately exponential burst light curve, and the distribution and composition of interstellar dust along the line of sight. Two of the most widely used interstellar grain models are the Mathis, Rumpl, \& Nordsieck (1977; MRN) model and the Weingartner \& Draine (2001; WD01) model. The MRN model is composed of silicate and graphite grains with a size distribution of $n(a) \propto a^{-3.5}$, which reproduces the observed extinction of starlight. The WD01 grain model additionally accounts for the observed infrared and microwave emission from the diffuse ISM by including sufficient small carbonaceous grains. In this paper we use the WD01 interstellar grain model and we assume that the dust is distributed uniformly along the line of sight.

The structure of this paper is as follows: We begin by describing the \chan~observation that was used in this analysis and we briefly discuss pile-up (\S~2). In \S~3 we present two analytic equations from Draine (2003) describing the scattering of X-rays from interstellar grains and the resulting halos, from which we derive an equation describing the predicted counts spectrum for an arbitrary range of angles and energies. We also discuss our treatment of the PSF. The derived equation is used to estimate the optical depth to GX 13+1 assuming uniformly distributed dust along the line of sight by fitting it to the observed surface brightness distribution (SBD). In \S~4 we demonstrate our method for calculating the expected delayed halo light curve of singly-scattered photons from a burst. In \S~5 we show that, given a particular set of assumptions, the 1--5 keV energy range and the 10--50\arcsec~annular region yield the highest time-averaged $S/N$ for viewing bursts in the halo. With this binning, simulations are carried out to test the feasibility of applying the TS73 method to type I X-ray bursts. We finish in \S~6 by summarizing our results. 

\section{Observations}
\begin{figure}
\centering
\includegraphics[width=3.5in]{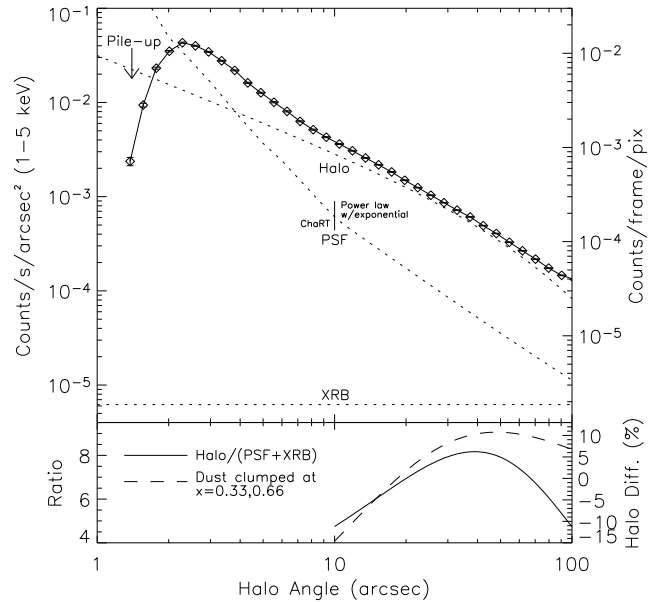}
\caption{\label{plgx} Surface brightness distribution of GX 13+1 (diamonds with error bars). The halo profile assumes uniformly distributed dust. The best-fit column density for 10--100\arcsec~is $N_{\rm H} = (2.75 \pm 0.06)  \times 10^{22}$ cm$^{-2}$, corresponding to $\tau_{\rm sca}(1~{\rm keV}) \approx 1.4$ using the WD01 grain model. For halo angles less than 10\arcsec~the PSF is modeled using ChaRT, and for halo angles greater than 10\arcsec~the PSF is modeled using a power law with an exponential term, using coefficients from Gaetz (2004). The pile-up fraction reaches 10\% at about 0.012 counts/frametime/pix (Smith et al. 2002), corresponding to $\theta_{\rm h} \la 3\arcsec$. The lower panel shows the ratio of the halo flux to the sum of the PSF and background fluxes ({\em solid curve}), and the fractional difference in the halo contribution to the SBD assuming half of the dust is uniformly distributed, and half is divided into two clumps at $x=0.33$ and $x=0.66$ ({\em dashed curve}).}
 
\end{figure}

In this paper, we use a 29 ks \chan~observation of GX 13+1 (Obs. ID 2708). The observation uses the Advanced CCD Imaging Spectrometer (ACIS), with the source focused on the back-illuminated ACIS-S3 chip, and with the High Energy Transmission Grating (HETG) inserted in the optical path. GX 13+1 has a substantial halo (Smith et al. 2002), yet it is not a consistent type I burster (two bursts were observed in Sept. 1989; Matsuba et al. 1995). Nevertheless, this observation allows us to test the potential for using bursts as distance indicators. We accomplish this by measuring the GX 13+1 SBD, from which we predict the size of a delayed burst signal in the halo as a function of delay time. Figure \ref{plgx} shows the observed surface brightness distribution of GX 13+1 (diamonds with error bars) and the best-fit halo (see \S~3). Statistical errors for the SBD are slightly less than 1\% for $\theta_{\rm h} \ga 10\arcsec$. Included in the figure is the contribution from the X-ray background (XRB) and the PSF, the treatment of which is in \S~3. Note the clear turnover in the observed SBD within the innermost 2--3$\arcsec$, resulting from the rejection of events due to pile-up. Pile-up is defined to be when more than one photon arrives in the same $3\times3$ pixel island in a single frame time. This can result in spectral hardening, ``grade migration," or rejected events in the case of pulse saturation. According to the \chan~Proposers' Observatory Guide v8.0 (CPOG), a pile-up fraction\footnote{The ratio of the number of detected events that consist of more than one photon to the total number of detected events.} of 10\% will very likely impact the observation. Smith et al. (2002) found the pile-up fraction reaches 10\% when the count rate per pixel is 0.012 counts/frametime/pixel, which corresponds to halo angles less than about 3\arcsec~in this observation. While pile-up in the GX 13+1 observation is most prevalent for halo angles $\la$ 5\arcsec, it likely affects the data to larger angles. Given the relatively large calibration uncertainties of the PSF (CPOG) and its increasing contribution nearer to the point source, we conservatively ignored halo angles less than 10\arcsec~in our final analysis.

Spectral response matrices and auxiliary response files for GX 13+1 were created with the standard CIAO scripts.\footnote{see http://cxc.harvard.edu/ciao/} The GX 13+1 source spectrum was fitted using XSPEC v11.3.2 with an absorbed blackbody plus a power law model, which allowed us to infer its hydrogen column density, measure the source flux, and predict X-ray halo spectra. The best-fit spectral parameters for this observation are $kT_{\rm bb} = 0.81 \pm 0.02$ keV, $\Gamma = 0.81^{+0.39}_{-0.52}$, and $N_{\rm H} = (2.46 \pm 0.04) \times 10^{22}$ cm$^{-2}$. 
\section{Grain-Scattered X-ray Halos and the PSF}
The theory of X-ray scattering and the resultant X-ray halos has been presented in detail in many previous publications (e.g., Overbeck 1965; Mauche \& Gorenstein 1986; Klose 1991; Mathis \& Lee 1991; Smith \& Dwek 1998; Draine 2003). In this analysis, we use two analytic equations of Draine (2003, eqs. [1] and [2] below), we assume uniformly distributed dust, and we apply the WD01 interstellar grain model\footnote{see http://www.astro.princeton.edu/$\sim$draine/dust/dust.html} with $R_{V} = 3.1$. Using the WD01 grain model, Draine (2003) obtained an approximation to the differential scattering cross section that applies to photon energies greater than 0.5 keV:
\begin{equation} \label{dsdw}
\frac{d\sigma(\phi,E)}{d\Omega} \approx \frac{\sigma_{\rm sca}(E)}
{\pi \phi_{\rm sca,50}^2 [1+(\phi/\phi_{\rm sca,50})^2]^2},
\end{equation} 
where $\phi_{\rm sca,50} \approx 360\arcsec/E({\rm keV})$ is the median scattering angle, which encloses half of the halo flux. This type of dependence on scattering angle $\phi$ implies that higher energy photons lead to halos that are more centrally peaked, and also implies that a greater fraction of the X-ray halo at higher photon energies results from smaller angle scatterings occurring close to the observer. This has observable implications in the delayed light curve following a type I burst (\S~4). The fraction of the total single-scattered halo flux interior to halo angle $\theta_{\rm h}$ is characterized by the normalized (dimensionless) dust density $\tilde{\rho} (x)$:
\begin{equation} \label{gh}
g(\theta_{\rm h}) \approx \int^1_0 \tilde{\rho}(x)\left[1+(1-x)^2\left(\frac{\phi_{\rm s,50}}
{\theta_{\rm h}}\right)^2\right]^{-1} dx,
\end{equation}
where $x$ is the fractional distance to the source. Using these relations, we define the fractional halo density (in units of arcsec$^{-2}$) as
\begin{equation} \label{grho}
g_{\rho}(E,\theta) \equiv \lim_{\delta \theta \to 0}
\frac{g(E,\theta+\frac{\delta \theta}{2})-g(E,\theta-\frac{\delta \theta}{2})}{2\pi \theta \delta \theta}.
\end{equation}
With an appropriate normalization, the fractional halo density for a given energy will produce approximately the same halo profile as $I_{\rm sca}(\theta)$, the conventional formula for halo surface brightness distributions (e.g., Mathis \& Lee 1991). With its explicit energy and halo angle dependence, however, this new quantity is particularly useful for calculating the delayed burst light curve in the halo.
\begin{figure}
\centering
\includegraphics[width=3.35in]{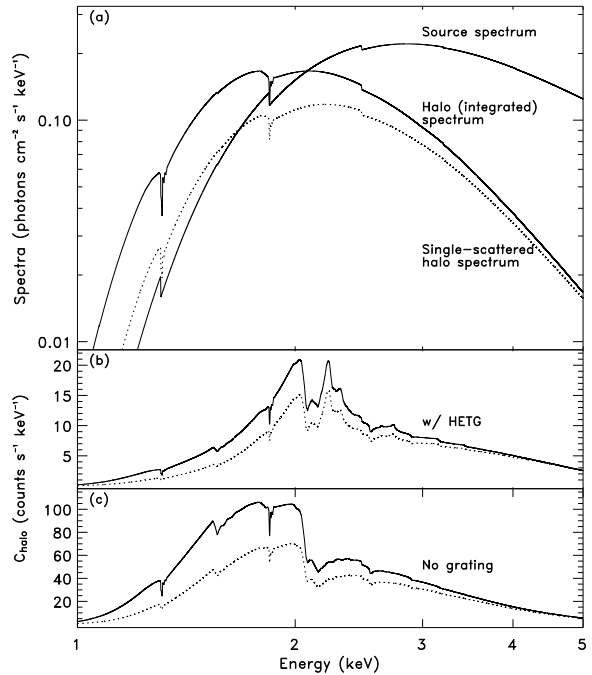}
\caption{\label{3pan} ({\em a}) Best-fit spectral model for GX 13+1 ({\em bold solid curve}), the halo spectrum integrated over all halo angles ({\em solid curve}), and the single-scattered halo spectrum ({\em dotted curve}), using the WD01 interstellar grain model. ({\em b})-({\em c}) The predicted counts spectra, obtained by multiplying the halo spectra by the {\em Chandra} zeroth-order effective areas for grating and non-grating observations. The decrease in effective area with off-axis angle has been ignored. For $\theta_{\rm h} \la 100\arcsec$ and $E \la 5$ keV, the decrease in area is only $\sim$1\%.}
\end{figure}
The spectrum of the X-ray halo integrated over all halo angles is a function of the source's spectrum prior to absorption and scattering along the line of sight $F_{\star}(E)$, the measured point source spectrum $F_{X}(E)$, and the optical depths through:
\begin{eqnarray} \label{s_halo}
F_{\rm halo}(E) &=& F_{\star}(E) e^{-\tau_{\rm abs}(E)} \left[1-e^{-\tau_{\rm sca}(E)}\right]  \\ &=& F_{X}(E)\left[ e^{\tau_{\rm sca}(E)} -1 \right], \nonumber
\end{eqnarray}
where $\tau_{\rm abs (sca)}$ is the optical depth to absorption (scattering) and $F_{\star} e^{-\tau_{\rm abs}}= F_{X} + F_{\rm halo}$. As discussed below, the distribution function for delayed photons scattering along the line of sight has a simple analytic form only for singly-scattered photons, yet the probability of a photon emerging in the halo after undergoing exactly $N$ scatterings is $\tau_{\rm sca}^N/N!$ relative to the observed source strength, so multiple scattering is likely important for sources with appreciable halos (Mathis \& Lee 1991). Nevertheless, singly-scattered photons generally appear earlier in the halo (after a shorter time delay) than photons that have been scattered multiple times (see Fig. 9, Alcock \& Hatchett 1978), so the effects of multiple scattering can be minimized by choosing small delay times. The halo spectrum for singly-scattered photons is obtained by replacing the term $\left[e^{\tau_{\rm sca}}-1\right]$ on the right side of eq. (\ref{s_halo}) with $\tau_{\rm sca}$. Figure \ref{3pan}{\em a} shows the measured point source spectrum, the inferred halo spectrum, and the single-scattered halo spectrum for GX 13+1. For an arbitrary energy range, the SBD as observed by \chan~(minus the PSF contribution) can be obtained from
{\small
\begin{eqnarray} \label{nhalo}
C_{\rm halo}({\Delta E, \theta}) &=&  \int\limits_{E_{1}}^{E_{2}} F_{X}(E) \left[e^{\tau_{\rm sca}(E)}-1\right] \\
& & \times g_{\rho}(E,\theta)  A_{\rm eff} (E,\theta) dE. \nonumber
\end{eqnarray}
}
%
%
By removing the fractional halo density from eq. (\ref{nhalo}), equivalent to integrating over all halo angles, the integrated counts spectrum is obtained (Fig. 2{\em b} and Fig. 2{\em c}). 

To accurately model the X-ray halo, it is extremely important to model the PSF so that the relative contributions of the dust-scattered and mirror-scattered halos can be separated. The \chan~Ray Tracer\footnote{see http://cxc.harvard.edu/chart/threads/index.html} (ChaRT) can be used to obtain a ray-traced PSF for any point on the detector and can be tailored specifically to the source spectrum; but while the simulated PSF represents the core of the source region well, it systematically underestimates the PSF at larger halo angles. We therefore chose to use the ray-traced PSF for halo angles less than 10\arcsec, normalizing the ray-traced PSF to the GX 13+1 source flux in 100 eV bands. Following Gaetz (2004), the {\em Chandra} PSF flux for halo angles greater than 10\arcsec~can be approximated by a power law with an exponential cutoff. Over a given energy range, the PSF can be approximated by
{\small 
\begin{eqnarray} \label{fpsf}
C_{\rm PSF}(\Delta E, \theta) &\approx& \int\limits_{E_{1}}^{E_{2}} \frac{F_{X}(E) \left[1-f_{\rm EEF,10}(E)\right]}{f_{10}\left(E,\frac{-\gamma (E)}{c(E)}\right)}\left[\frac{\theta}{10}\right]^{-\gamma (E)}  \\
& & \times e^{-c(E) \theta} dE,\nonumber
\end{eqnarray}
}
where $f_{\rm EEF,10}(E)$ is the encircled energy fraction at 10\arcsec~and is given in the CPOG for various energies, and the denominator is defined as 
\begin{equation}
f_{10}(E,\theta_{\rm h}) \equiv \int\limits_{10}^{\theta_{\rm h}} 2 \pi \theta \left[\frac{\theta}{10}\right]^{-\gamma (E)} e^{-c(E) \theta} d\theta,
\end{equation}
where the units of angle are arcsec. The denominator of eq. (\ref{fpsf}) is simply a normalization; $\frac{-\gamma (E)}{c(E)}$ is the inflection point of the power law times exponential function. The coefficients $\gamma (E)$ and $c (E)$ were measured in 1 keV bands by Gaetz (2004) using observations of Her X-1. Integration was performed by interpolating the values of $f_{\rm EEF,10}(E)$, $\gamma (E)$, and $c (E)$ to form a nearly continuous grid. 

The scattering optical depth to GX 13+1 was found by matching the observed SBD to that produced by the sum of eqs. (\ref{nhalo}) and (\ref{fpsf}) and the measured XRB (see Fig. \ref {plgx}). Note, however, that the use of eq. (\ref{nhalo}) assumes photons scattered two or more times produce the same halo SBD as singly-scattered photons, although multiply-scattered photons tend to produce somewhat broader halo profiles (Mathis \& Lee 1991). The best-fit hydrogen column density leading to the observed halo is $(2.75 \pm 0.06) \times 10^{22}$ cm$^{-2}$ (assuming a 20\% [10\%] uncertainty in the PSF [XRB] contribution), corresponding to $\tau_{\rm sca}(1~{\rm keV}) \approx 1.4$ using the WD01 interstellar grain model, while the best-fit column density using the a blackbody plus power law model is $N_{\rm H} = (2.46 \pm 0.04) \times 10^{22}$ cm$^{-2}$ (\S~2). The 8--14\% difference could be attributed to an incorrect assumption of uniformly distributed dust, to an inaccurate grain model, to the assumption that the SBDs for singly- and multiply-scattered photons are the same, or perhaps the gas-to-dust ratio along the sightline differs from the Galactic mean. 

Finally, the measured halo SBD can also be used to constrain the distribution of dust along the line of sight by applying eq. (\ref{nhalo}) for various dust distributions. In the bottom panel of Fig. \ref{plgx} we show the fractional difference in the halo contribution to the SBD assuming half of the dust is uniformly distributed, and half is divided into two clumps at $x=0.33$ and $x=0.66$ (approximated as Gaussians with $\sigma=0.1$). The statistical errors in the SBD are much smaller than the fractional difference due to different distributions. 

\section{Delayed Burst Light Curve in the Halo}
In order to be able to clearly observe the delayed pulse of radiation from type I X-ray bursts in the X-ray halo, the product of the source flux and scattering optical depth must be large enough to provide adequate statistics despite smearing of the delayed burst photons over many kiloseconds. A burst light curve is shown by Thompson et al. (2005, see Fig. 1) for Ginga (GS) 1826-238, and can be well-approximated by an exponential with $e$-folding time $t_{\rm b} \approx 60$ s. In general, burst decays are shorter at higher photon energies due to a softening of the burst spectrum during the decay, but we treat $t_{\rm b}$ as energy-independent for simplicity. The 60 s burst scale time for GS 1826-238 is relatively long; in fact, the majority of type I bursters have scale times less than 20 s (see Fig. 4, Cornelisse et al. 2003). Unfortunately, the amount of dust towards GS 1826-238 is too low to apply the TS73 method, with $N_{\rm H} \approx 1.9 \times 10^{21}$ cm$^{-2}$ (Dickey \& Lockman 1990) implying $\tau_{\rm sca} \sim 0.05$ at 1 keV (Predehl \& Schmitt 1995). 

The small size of the expected signal from a type I X-ray burst is immediately apparent considering that approximately $(F_{\rm peak}-F_{\rm persistent}) \tau_{\rm sca} t_{b}$ singly-scattered photons cm$^{-2}$ keV$^{-1}$ are expected over all halo angles and delay times for an X-ray burst approximated by an exponential. At halo angles larger than 100\arcsec, the burst photons are smeared to such a large extent that any signature of the delayed burst is hard to observe. To begin, it is straightforward to show that the time delay of photons scattered at a fractional distance $x$ to the source is
\begin{equation} \label{td}
t_{\rm d}=\frac{D\theta^2}{2c} \frac{x}{1-x} \approx 1.21 \frac{D}{\rm kpc} \left(\frac{\theta}{\rm arcsec}\right)^{2}
\frac{x}{1-x}~{\rm s}.
\end{equation}
We have invoked $\phi \approx \theta/(1-x)$, which is true for small angles and applies to all X-ray scattering. Although the time delay for a photon scattered at a known fractional distance and angle is simple to derive, in practice we have to integrate over a distribution function of time-delayed photons, which in the single scattering approximation is 
\begin{equation} \label{H}
H(t_{\rm d},\theta,E)=\frac{\tilde{\rho}(x)}{t_{\rm d}+ t_{\theta}} \frac{d\sigma(\phi,E)}{d\Omega},
\end{equation}
where we have defined $t_{\theta} \equiv D \theta^2 / 2 c$. This distribution can be understood by noting that the delayed light curve in the halo is only a function of the time delay and viewing angle, although physically, the light curve is governed by the distribution of dust, and the differential scattering cross section of the dust; the latter are functions of $x$ and $\phi$, respectively. Therefore, $H(t_{\rm d},\theta)$ is simply the product of $\tilde{\rho}(x)$, $d\sigma(\phi)/d\Omega$, and the Jacobian determinant relating the area elements $\vline \partial(x,\phi)/\partial(t_{\rm d},\theta) \vline$ (TS73). Although $d\sigma/d\Omega$ is a function of the scattering angle, it is implicitly a function of the time delay through $\phi \approx \theta/ (1-x)$. An alternative derivation of $H(t_{\rm d},\theta)$ is provided by Alcock \& Hatchett (1978, see eq. [17]), who also derived formulae describing the propagation of a pulse of radiation in the large optical depth limit.

The distribution of single-scattered delayed photons has a form similar to a decaying exponential, falling off more slowly at larger halo angles and for larger source distances. Considering that more than half of the space along the line of sight is covered within about $600(D/{\rm 5~kpc})(\theta_{\rm h}/10\arcsec)^2$ s ($dx/dt_{\rm d} \sim t_{\rm d}^{-2}$), this type of functional behavior is not surprising. The distribution of delayed photons also shows that the extent of smearing increases dramatically with halo angle, which can also be seen in eq. (\ref{td}), where $t_{\rm d} \propto \theta^{2}$. While the delayed light curve of a $\delta$-function burst at a given angle can be obtained directly from $H(t_{\rm d},\theta)$, for an arbitrary light curve one must convolve the non-delayed light curve $I(t)$ and the distribution of time-delayed photons:
\begin{equation} \label{B}
\tilde{B}(t_{\rm d},E,\theta)=\frac{\int\limits_{-\infty}^{t_{\rm d}} 
I(t)H(t_{\rm d}-t,E,\theta) dt}
{\int\limits^{t_{\rm d}}_{-\infty} H(t_{\rm d}-t,E,\theta) dt}.
\end{equation}
The delayed burst light curve (in units of counts/s) for singly-scattered photons for a given range of halo angles and energies, and for a given dust distribution and source distance is
{\small
\begin{eqnarray} \label{bo}
B(t_{\rm d}, \Delta E,\Delta \theta)&=&\int\limits_{E_{1}}^{E_{2}} \int\limits_{\theta_{1}}^{\theta_{2}}
\tilde{B} (t_{\rm d}, E,\theta) \langle S_{\rm b}(E) \rangle \left [ e^{\tau (E)}-1 \right ]  \\ & & \times A_{\rm eff} (E,\theta) g_{\rho}(E,\theta) 
2 \pi \theta d\theta dE, \nonumber
\end{eqnarray}
where $\langle F_{X,{\rm b}}(E) \rangle$ is the average burst spectrum (with the persistent spectrum subtracted out), and requiring eq. (\ref{B}) to be normalized, which is obtained naturally if $I(t)$ is normalized, i.e. $I(t) = t_{\rm b}^{-1}\exp{(-t/t_{\rm b})}$, for a type I burst approximated by an exponential. Figure \ref{blc} shows the normalized delayed light curves for X-ray bursts with $t_{\rm b} = 60$ s and $E = 2$ keV for sources at three distances with uniformly distributed dust. The ratio of the peak to persistent flux was chosen to be 7, similar to that of GS 1826-238. This ratio is fairly typical of brighter bursters (low persistent emission sources have much higher ratios; Cornelisse et al 2002), although the typical range is rather wide. To facilitate interpretation of Fig. \ref{blc}, consider the case of a unit annulus centered on 5\arcsec~and a 4 kpc source distance ({\em top panel, dotted curve}): About 12\% of the total burst fluence would be expected in a 100 s bin centered at a time delay of 100 s. Small angles are shown, though the general trend of a smaller signal at larger angles due to greater smearing is clear.

\section{Simulations}
\subsection{Maximizing the Signal-to-Noise Ratio}
The feasibility of using type I bursts as distance indicators was tested by simulating burst light curves in the halo, including both statistical and systematic errors, and finding the deviation of the best-fit distances. Systematic errors include uncertainties in the calibration of the PSF, and in the measurements of the optical depth, the peak and persistent fluxes, and the scale time of the burst. Due to the small size of the expected delayed burst signal, it is essential to maximize the $S/N$ ratio (where the signal refers to counts due to the burst and the noise refers to the total Poisson noise) by binning the data in a favorable configuration for the simulations. At higher energies, for example, the dust-scattered halo is rapidly decreasing while the PSF contribution is increasing. Not only does this binning depend on the spectrum and optical depth of the source, the observing telescope's effective area and PSF, and the distribution and composition of interstellar dust, but also on the time after the onset of the burst. Due to the energy dependence of the differential scattering cross section, harder photons appear at smaller time delays and halo angles relative to softer photons. Our goal was therefore to find an effective binning of the data averaged throughout the dynamical time period of the delayed burst evolution, meaning the period when the delayed light curves for sources at different distances differ by more than a few percent. For uniformly distributed dust, the dynamical period is isolated to time delays less than $\sim$10 ks because the delayed burst light curves for sources at different distances converge at longer time delays (see Fig. \ref{blc}). 
\begin{figure}
\centering
\includegraphics[width=3in]{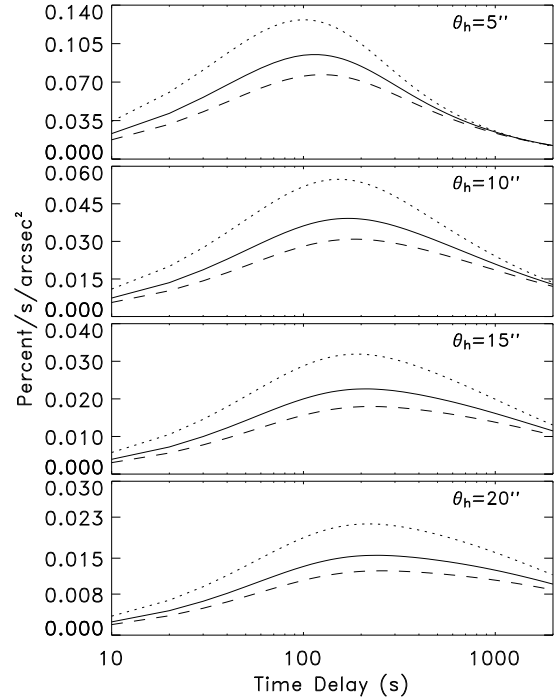}
\caption{\label{blc} Predicted delayed burst light curves for small halo angles ({\em top} to {\em bottom}: $\theta_{\rm h}=[5,10,15,20]$ arcsec), and for three source distances ({\em dotted}: 4 kpc, {\em solid}: 6 kpc, {\em dashed}: 8 kpc) assuming uniformly distributed dust, 2 keV photons, and $t_{\rm b} = 60$ s. The $y$-axis units are the percent of total burst photons appearing per second per unit solid angle.}
\end{figure}
The time-averaged optimal binning was obtained by selecting an energy grid from 1-6 keV with 0.5 keV bins, and a radial grid from 10--160\arcsec\footnote{Halo angles larger than 160\arcsec~were not analyzed due to the presence of a diffuse ``transfer swath'' caused by the diffracted spectrum, analogous to the so-called ``transfer streak'' centered on the source.} with 10\arcsec~bins. Because the cross section for X-rays scattering off of interstellar grains is proportional to $\sim$$E^{-2}$, and because the smearing of the burst signal increases dramatically with angle ($t_{\rm d} \propto \theta^{2}$, see Fig. \ref{blc}), the lower bounds of the trial binning were fixed to 1 keV and 10\arcsec. Using 200 s time bins, trial combinations of angle and energy were selected. The included photon energies were increased incrementally, and for each energy increment the included range of halo angles was also increased incrementally. For each trial binning, the time-averaged $S/N$ ratio was calculated. We find that averaged throughout the dynamical portion of the delayed burst, the optimal annular region is about 10--50\arcsec~and the optimal energy is from about 1--5 keV for a source with $\tau_{\rm sca}(1~{\rm keV}) \approx 1.4$ and the spectrum of GX 13+1 observed with \chan's effective area and PSF. This range of halo angles is not surprising considering the ratio plot in the lower panel of Fig. \ref{plgx}, although the reader should note that we did not optimize the plotted ratio but rather the $S/N$ ratio.

\subsection{Distance Measurement Method and Potential Accuracy}
The best-fit distance to the simulated light curves was measured using $\chi^2$-minimization. A five-dimensional array composed of the predicted flux as a function of delay time for an arbitrary distance, halo angle, and energy was created. The energy and angle degrees of freedom were removed by integrating over the predicted single-scattered burst halo spectrum for GX 13+1, and the angles 10--50\arcsec~assuming an azimuthally-symmetric halo profile. The persistent halo was treated as a constant. The value of $\chi^2$ was then calculated by comparing predicted light curves for sources at distances in 0.01 kpc increments to each trial of simulated data; the best-fit distance for each simulated light curve had the smallest fit statistic. One thousand trial light curves were used in each simulation. 
\begin{figure}
\centering
\includegraphics[width=3.5in]{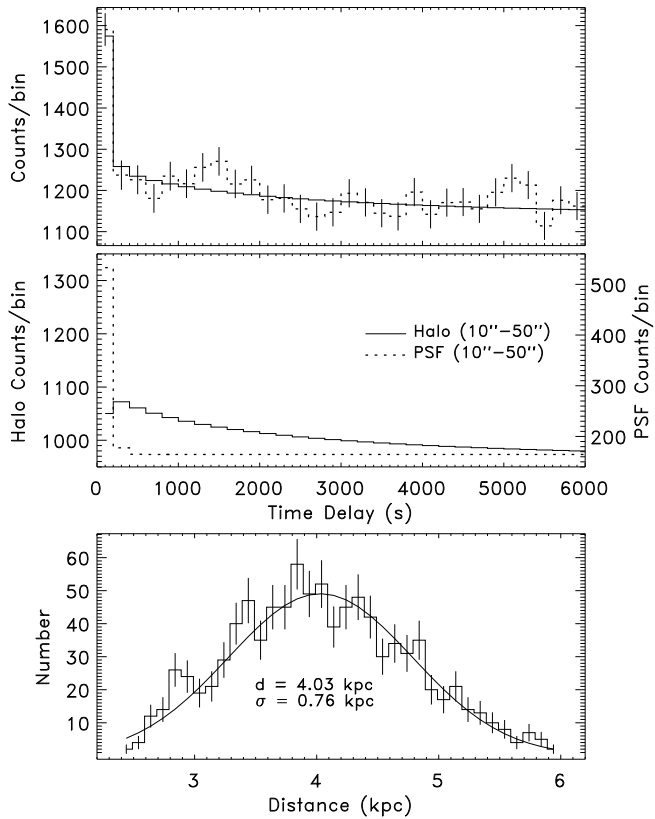}
\caption{\label{sim} {\em Top:} Predicted 10--50\arcsec~halo light curve and a sample simulated light curve after the onset of a type I X-ray burst for a source at 4 kpc, assuming only statistical errors. {\em Middle:} Relative contributions to the predicted light curve. {\em Bottom:} Histogram (0.1 kpc bins) of the best-fit distances for one thousand simulated light curves, and the Gaussian fit to the distribution.}
\end{figure}
We first examined the potential accuracy of a distance measurement using a \chan~grating observation assuming the only uncertainties are due to Poisson noise. By simulating burst light curves for source distances from 3 kpc to 11 kpc in 0.5 kpc increments, we find that as the source distance becomes larger, the distribution of best-fit distances becomes increasingly distorted from a Gaussian due to a slight tail for larger source distances (forming a Gamma-like probability density function). This makes sense because the differences in the delayed light curves become smaller at larger distances. Therefore, a negative statistical error (fewer counts than the average) will overestimate the distance more than an equal positive statistical error will underestimate the distance. Figure \ref{sim} shows predicted and simulated 10--50\arcsec~halo light curves, the relative contributions of the halo and the PSF, and a histogram of the best-fit distances for a source distance of 4 kpc. We find the fractional error on the inferred distance to a source 4 kpc away is $^{+22}_{-17}$\% for a single burst (68\% confidence). At 6 kpc and 8 kpc, the errors are $^{+29}_{-19}$\% and $^{+40}_{-21}$\% respectively. The positive and negative errors increase approximately linearly with distance.  
For a source with a smaller optical depth to scattering than GX 13+1, e.g. $\tau_{\rm sca} (1~{\rm keV}) \approx 0.5$, roughly five aligned and summed bursts would be required to obtain similar accuracy. By including realistic systematic uncertainties of 10\% for the scattering optical depth, 20\% for the PSF contribution, and 5\% on the flux and scale time of the burst, the potential accuracy decreases by only $\sim$10\% for distances between 3 kpc and 11 kpc. Future large area observatories with adequate spatial resolution (e.g., {\em XEUS} and {\em Constellation-X}) will effectively minimize the Poisson noise, in which case the distance measurement accuracy is limited by systematics. By simulating data with systematic errors only (which is also equivalent to stacking many bursts), the uncertainty in the distance measurement is 7\% at 3 kpc, increasing approximately linearly with distance to 12\% at 11 kpc . 

The simulations above assumed a constant persistent halo, but due to the nature of the accretion processes in LMXBs, there is likely to be source variability on all time scales. With reliable flux measurements (from a grating observation), handling the PSF is straightforward because its contribution is essentially a constant fraction of the point source flux (see eq. [\ref{fpsf}]), and there is no time delay associated with it. Likewise, halo variability can be handled by including additional terms in the light curve $I(t)$ in eq. (\ref{B}), and appropriately changing the normalization. As a worst case scenario, however, assume that variability exists on 200 s scales, yet we still assume the persistent halo is constant in the calculation of the delayed light curve. To test the effect of this assumption, we calculated delayed light curves for halo angles of 10\arcsec~to 50\arcsec~in 10\arcsec~increments for time-averaged constant source light curves (on a scale much longer than 200 s) but with stochastic variability on 200 s scales with rms amplitudes of 10\%, 20\%, 30\%, and 40\%. Because this variability is spread out over many kiloseconds due to scattering along the line of sight, the rms amplitude is smaller in the halo. Interestingly, the halo rms amplitudes seem to be only weakly dependent on the halo angle, with corresponding rms amplitudes are $\sim$2\%, 4\%, 5--7\%, and 6--9\% respectively. These percentages were added to the statistical and systematic errors for the persistent halo contribution in our simulations. For the worst case of 40\% point source variability, we find the inferred distance errors increases by about 15\% for a grating observation, e.g., for a 6 kpc source distance and a simulation including systematics, the error increases from about 30\% to 35\%. This can also account for dust distributions that are overall uniform, but patchy on scales $x \ll 1$.

A non-grating observation with \chan~increases the effective area and signal by about a factor of 4 (see Fig. 2), but the reduction in Poisson noise is offset by larger systematic uncertainties in the flux measurements and the burst scale time. Without the dispersed grating counts, one is required to extract spectral and flux characteristics from the source transfer streak (e.g., Smith et al. 2002, Clark 2004), generally a much coarser and poorly calibrated method. More importantly perhaps, a non-grating observation would significantly increase the extent of pile-up, forcing one to look at larger halo angles where the evolution of the delayed burst occurs on longer time scales.

\section{Conclusions}
In this paper, we studied the applicability of the TS73 method to the delayed signature of type I X-ray bursts in the halo. The signal produced by an X-ray burst in the halo is invariably small. Regular type I bursts last $\sim$5--150 s, but the scattering geometry, which governs the distribution of time delays, assures that these photons will be spread out over time delays greater than $\sim$10 ks for uniformly distributed dust. Nevertheless, if the physical parameters governing the light curve of the burst in the halo are well-known, the statistics provided by \chan's effective area can yield a distance measurement to about 25\% accuracy with a single burst. By stacking many bursts, systematic errors likely limit the potential accuracy to roughly 10\%. Although non-grating observations have improved statistics due to larger effective area, pile-up precludes accurate flux measurements (although the transfer streak can be used) and require the use of larger halo angles, decreasing the $S/N$ as the evolution of the delayed burst light curve occurs on longer time scales. Source variability on 200 s scales with 40\% rms amplitude increase the distance uncertainty by $\sim$15\%. The quoted errors assume very favorable source characteristics: (1) a bright source with a  flux $\sim$0.6 photons/cm$^{2}$/s during persistent emission from 1--5 keV, (2) a large optical depth to scattering ($\tau_{\rm sca}\approx$1.4 at 1 keV), both of which increase the $S/N$ ratio of the delayed burst light curve in the halo, (3) a burst with large total fluence, characterized by $t_{\rm b} = 60$ s and $F_{\rm peak}/F_{\rm persistent} = 7$. In the general case, the uncertainty in the inferred distance will be somewhat larger. Finally, note that the interstellar grain model represents an additional source of systematic error.

In many cases the distance to Galactic X-ray sources has been estimated using other methods, e.g., using the requirement that the inferred peak luminosity is sub-Eddington, or using optical photometry of the (low-mass) binary companion, and the potential accuracy of the TS73 method applied to bursters may only provide additional distance constraints for a small fraction of all type I bursters. Moreover, if the distribution of dust differs significantly from the assumed or inferred distribution, the distance measurement can be unreliable. A future observatory with a larger effective area and arcsec resolution would provide superior statistics, however, making the TS73 method an attractive means of measuring the distance to a larger subset of the population of absorbed X-ray bursters. 

\acknowledgements
We acknowledge support from NASA grant NAS5-30702. We thank Peter Predehl for advice on modeling the PSF, and Bruce Draine for help interpreting the equations from Draine (2003). TT thanks Christel Smith for many helpful discussions. We also appreciate many comments and suggestions by the referee that significantly improved this paper. This research has made use of data obtained through the High Energy Astrophysics Science Archive Research Center Online Service, provided by the NASA/Goddard Space Flight Center. The analysis of the \chan~data made extensive use of the Chandra Interactive Analysis of Observations (CIAO), \url{http://cxc.harvard.edu/ciao}.

\end{document}